\documentclass[twocolumn, 10pt]{IEEEtran}

\usepackage{url}
\usepackage{hyperref} 
\usepackage{graphicx}
\usepackage{amssymb}
\usepackage{amsmath}
\usepackage{mathtools}
\usepackage{cite}
\usepackage{stfloats}
\usepackage{subfigure}
\usepackage{psfrag}
\usepackage[mathscr]{euscript}
\usepackage{acronym}  
\usepackage{booktabs} 
\usepackage[usenames,dvipsnames]{color}
\usepackage{graphicx}

\DeclareMathAlphabet{\mathrmbf}{\encodingdefault}{\rmdefault}{bx}{n}

\newcommand{\mcal}{\mathcal}
\newcommand{\mbb}{\mathbb}

\newcommand{\msf}{\mathsf}

\usepackage{array}
\newcolumntype{P}[1]{>{\vspace*{0.07cm}\centering\arraybackslash}p{#1}}
\newcolumntype{M}[1]{>{\vspace*{0.07cm}\centering\arraybackslash}m{#1}}

\usepackage{pbox}

\begin{document}
%

\title{Reconfigurable Security:\\ Edge Computing-based Framework for IoT}

%
%
%
%
%

\author{	
	\vspace{0.2cm}
	Ruei-Hau~Hsu, \textit{Member, IEEE}, 
        Jemin~Lee, \textit{Member, IEEE}, 
        Tony~Q.~S.~Quek, \textit{Senior Member, IEEE}, and
		Jyh-Cheng~Chen, \textit{Fellow, IEEE}
\IEEEcompsocitemizethanks{\IEEEcompsocthanksitem R.-H. Hsu was with iTrust, Centre for Research in Cyber Security, Singapore University of Technology and Design, Singapore, 487372. He is now with Data Storage Institute~(DSI), Agency for Science, Technology and Research~(A*STAR), Singapore, 138634.
(E-mail: richard\_hsu@sutd.edu.sg)
\IEEEcompsocthanksitem
J. Lee is with Department of Information and Communication Engineering, Daegu Gyeongbuk Institute of Science and Technology (DGIST), Korea, 43016.(Email: jmnlee@dgist.ac.kr)
\IEEEcompsocthanksitem 
T.~Q.S. Quek is with Information Systems Technology and Design Pillar, Singapore University of Technology and Design, Singapore, 487372. (Email: tonyquek@sutd.edu.sg)
\IEEEcompsocthanksitem
J.-C. Chen is with Department of Computer Science, National Chiao Tung University, Taiwan, 30010.
(Email: jcc@cs.nctu.edu.tw)
\IEEEcompsocthanksitem
The corresponding author is J. Lee. 
}}

\IEEEcompsoctitleabstractindextext{%
\begin{abstract}

In various scenarios, achieving security between IoT devices is challenging since the devices may have different dedicated communication standards, 
resource constraints as well as various applications. 
In this article, we first provide requirements and existing solutions for IoT security. 
We then introduce a new reconfigurable security framework based on edge computing, 
which utilizes a {near-user edge device}, i.e., \textit{security agent}, to simplify key management and offload the computational costs of security algorithms at IoT devices. 
This framework is designed to overcome the challenges including high computation costs, low flexibility in key management, and low compatibility in deploying new security algorithms in IoT, especially when adopting advanced cryptographic primitives.
We also provide the design principles of the reconfigurable security framework, the exemplary security protocols for anonymous authentication and secure data access control, and the performance analysis in terms of feasibility and usability. 
The reconfigurable security framework paves a new way to strength IoT security by edge computing.

\end{abstract}

}
\maketitle

\IEEEdisplaynotcompsoctitleabstractindextext

\IEEEpeerreviewmaketitle

\section{Introduction}\label{sec:introduction}

Due to the diverse types of consumer electronics, people's life has been changed dramatically. The devices are interconnected by diverse types of communication technologies to the Internet, known as {I}nternet of {T}hings~(IoT) to exchange information. Nowadays, {IoT devices} are widely deployed based on the existing standards for various applications such as smart home, smart city, body networks, smart grid, vehicular ad-hoc networks, and autonomous control systems. Many IoT alliances and consortia, e.g., Alljoyn, IEEE P2413, IPSO, OCF, OMA, etc., have proposed their own standard frameworks for IoT among the developed standards. They attempt to manage things, devices, the provided information, and their computing ability as resources, which can be interconnected and utilized. In this context, security is of the first priority to guarantee the availability and functionality of IoT applications. Only basic security protections, i.e., authenticated key exchange and access control, for the communications has been addressed in IoT related standard frameworks.
%


The security mechanism for IoT should support heterogeneous types of devices and communication standards for applications, which leads to the requirements of comprehensive security protections~\cite{RS_HYC11}, such as anonymous protection and fine-grained secure access control. Therefore, new security framework is required, which can simplify deployment of security solutions and minimize the change of existing systems.

\subsection{Security Issues in IoT related Applications}
The applications of IoT can be categorized into the following domains: Internet of transportation, Internet of senor/controller, Internet of energy, and device-to-device/machine-to-machine communications. Exemplary applications of these domains include vehicular ad-hoc networks~(VANETs), smart home, and smart grid. The security requirements of each application can be various according to application scenarios. For instance, VANETs need to keep the identity of each vehicle anonymously against location/session traceability, the smart home needs to protect the identity of each controller to conceal sensitive behaviors of each individual, and the smart grid will need to conceal electricity consumption against analysis in appliance usage. Besides the specific security requirements of each IoT application, there are generically challenging issues in guaranteeing security in IoT as follows.

\begin{itemize}
\item Key management of cross-application is complicated as each application may manage its security keys for specific security purposes~\cite{IoTSec_RSSS16}. A user device engaged in multiple applications needs to manage multiple security keys or passwords. This will increase the risk of key disclosure and endanger the security of services.
\item Advanced cryptographic algorithms, e.g., group signatures~(GS)~\cite{GS_BBS04} for anonymous communications and attribute-based encryption~(ABE) for access control on data, need to be performed on IoT resource-constrained devices. This will make real-time IoT application failed as the computation of the advanced algorithms cannot be completed in a pre-defined period.
\item Flexibility to support various security protections for diverse IoT applications with minimum changes on fundamentals, i.e., standard protocol, softwares, and hardwares, is a significant factor to developers for usability.
\end{itemize}
%

{Due to above issues, the challenges in satisfying IoT security requirements\footnote{The IoT security requirements can be divided into the two major categories, i.e., \textit{secure communication} and \textit{data security}.} include 
1) the \emph{complexity of key management} for diverse application and data models, 2) the \emph{computational infeasibility} of resource-constrained IoT devices for advanced cryptographic algorithms with high computation complexity, 3) the \emph{inflexibility in supporting new security functions} due to independent design of each communication standard and the layered architecture of networking. 

\subsection{Possibly Reconfigurable Solutions for IoT Security}
For the security issues in IoT, the following existing solutions can be considered.
\begin{enumerate}
\item {\em AAA Framework and Extensible Authentication Protocol:} 
Authentication, authorization, accounting~(AAA) framework supports security services for applications such as mobile IP, network access service~(NAS), and session initiation protocol~(SIP), in both local and roaming scenarios.\footnote{Remote authentication dial-in User Service~(RADIUS) and Diameter are two standard protocols adopting AAA frameworks,  where the latter is the newer standard but not fully backward compatible with RADIUS.}
The {EAP}~\cite{EAP_CW05} is a standard authentication framework {to support the flexibility of various security protocols based on AAA framework}, widely used in wireless networks.
AAA and EAP offer certain flexibility on key management and various security protections~\cite{EAP_FLH13}. However, the issue of computation costs in applying advanced algorithms still remains.
\item {\em Access Control Techniques:}
The role-based access control~(RBAC) has been widely used in software systems and applications for operating and managing resources. The RBAC is originated by using the concept of user/group to grant permissions to access files in UNIX system~\cite{RBAC_FSGKC01}. 
The attribute-based access control~(ABAC)~\cite{AttributeEnc_GSPW06}, different from RBAC, provides a fine-grained access control by controlling each access with the given policies and the attributes of users. Those access control techniques may offer the flexibility in adopting various security measures by regarding each security service as object.
\item {\em Single Sign On Mechanism:} The single sign on~(SSO) is an access control technique that makes a user logs into multiple software systems using one account. Many related standards have been proposed, such as the lightweight directory access protocol~(LDAP)~\cite{RFC4511_LDAP}, Kerberos network authentication service~\cite{RFC1510_Kerberos}, smart card based authentication, and security assertion markup language~(SAML). In SSO, user account information is stored in databases on authentication servers. For every user accessing to a software system, the SSO-based authentication involves the user, the authentication server, and the target system. SSO is akin to access control system and also facilitates the flexibility of providing security services.

\end{enumerate}

The above solutions allow certain flexibility to adopt comprehensive security measures and to control the access to security services. However, the key management remains as an issue since each device needs to maintain multiples security keys or certificates for diverse applications/services. Moreover, how to make the computation of advanced cryptographic algorithms feasible on the resource-constrained devices is still an issue. Hence, a framework for the requirements of IoT security is essential, which will be introduced for the above two issues in this article.}
%

\subsection{Reconfigurable Security with Edge Computing}\label{subsec:Rec_Sec}
The edge computing is a new computing model~\cite{EdgeCom_SCZLX16}, where a near-user device with stronger computing power provides required resources for the applications of other resource-limited IoT devices. Based on edge computing, the challenges of high computation costs, low flexibility, and incompatibility, in supporting security with advanced cryptography can be relieved by the new framework below.

We introduce a {new \emph{reconfigurable security} framework for IoT~(ReSIoT)} to overcome security challenges 
without changing the architectures or re-designing the standard protocol flows of IoT applications. The framework utilizes a new component, i.e., {\it security agent}~(SA), which is a near user-edge device such as wireless router, base station, service router, etc. The computation capability of SA is generally more powerful than most of resource-limited IoT devices and it can be used to offload the overhead of cryptographic computations at the resource-limited devices and centralized computing infrastructure. 

Through the reconfigurable security, each device involving a {security function} (e.g., authentication, access control, etc.) only needs to guarantee secure communication with the SA by maintaining keys with SA.
The SA generates and distributes required security information with its security key, registered to a global key management system~(GKMS), to {complete} corresponding security procedures between IoT devices. 
The benefits of the introduced reconfigurable security for IoT can be summarized as follows:
\begin{enumerate}
\item The key management can be simplified from application level to user level in terms of IoT devices.
\item Due to stronger computation capability of SA, even the low-end devices will be able to be protected by advanced security algorithms requiring high computation costs. 
\item Different from the previous concept of reconfigurable security in 3G/4G networks~\cite{RS_AMC02,RS_HYC11},\footnote{The concept of reconfigurable security has also been proposed for 3G/4G networks~\cite{RS_AMC02,RS_HYC11} to resolve the security of 3G/4G interworking and roaming mobile devices. However, these frameworks only address requirements of secure communications, and also adopting the frameworks needs to change original system architectures. } 
the introduced reconfigurable security for IoT does not require to change
system architectures of the original protocols and standards, and also considers all possible security requirements that can be fulfilled by cryptographic countermeasures.
\item Compared to cloud-based solutions, exploiting near-user edge devices for IoT security provides better scalability and usability. 
\item Even cloud-based computing may also help to offload computation overhead of security protection, users may need to give their security keys to cloud servers, while the reconfigurable security remains security keys concealed by users.
\end{enumerate}
{This article} gives light on the potential of reconfigurable security as an essential security {framework} for IoT, and organized as outlined below. 
In Section II, security requirements of IoT is presented, and the potential security solutions and their limitations are discussed in Section III. 
In Sections IV and V, {the system design and constructions of reconfigurable security functions are provided,} and {their performances are evaluated in Section VI}. 


\section{Security Requirements of ReSIoT}\label{sec:sec_requirements}

\begin{figure*} [!t]
	\begin{center}
		\includegraphics[height=9.0cm]{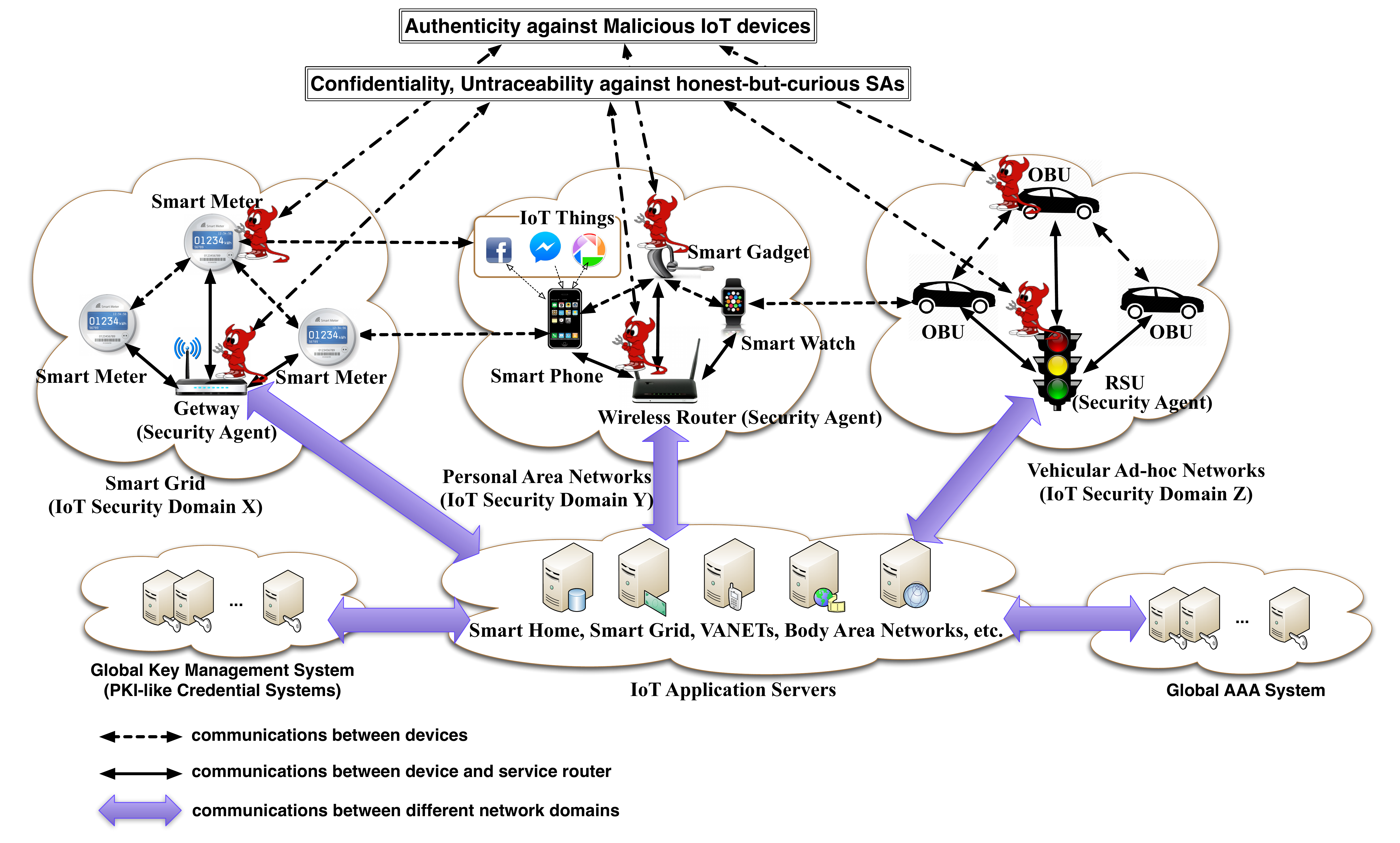} \caption{The system architecture and adversary model of IoT reconfigurable security.}
		\label{fig:SA_system_model}
	\end{center}
	\vspace{-0.3cm}
\end{figure*}
\subsection{Security Architecture}

Besides of the common security requirements of confidentiality, integrity, availability, and non-repudiation, the ReSIoT gives rise to additional security issues regarding the computation of security functions~(SFs) on SAs for IoT devices. The security requirements specially aim at two kinds of adversaries, malicious IoT devices and honest-but-curious SAs, as shown in Fig.~\ref{fig:SA_system_model}. A malicious IoT device may misbehave during the interactions with the other IoT devices. A honest-but-curious SA may intercept the exchanged messages and trace the communication footprints among IoT devices. Hence, a reconfigurable security function~(RSF) constructed by a specific SF should fulfill the additional security requirements as follows.
	%
%
\begin{itemize}
\item{\bf Confidentiality against honest-but-curious SA:} The confidentiality of message exchanges between IoT devices should be guaranteed even SA helps on computing SFs involving the exchanged messages. The SA should not learn any information from the communications between IoT devices and the procedures of performing RSFs.
\item{\bf Authenticity against malicious IoT devices:} The identity of each IoT device should be verifiable by the belonging SA. Before computing the specified SF for the requesting IoT devices, SAs should authenticate the devices first. This guarantees that unauthorized IoT devices do not abuse the capability of computing SFs by SAs.
\item{\bf Untraceability to IoT devices against honest-but-curious SA:} Every SA should not be able to trace the identity of each IoT device in a communication session launched by IoT devices even if the SA helps to compute the SF of the session. This guarantees that footprints of all communications are kept secret to SAs involved in the computation.
\end{itemize}
By achieving the above three security requirements, SAs can authenticate each IoT device before computing the specified SF for them. The confidential messages of IoT devices cannot be exposed to SAs. Moreover, the identity of devices and communication sessions are unlinkable, so SAs cannot trace the identity of any devices during all communication sessions between IoT devices.
\section{System Architecture of ReSIoT}\label{sec:proposed_IoT_RS}
In this section, we first introduce the system architecture, including the proposed SA, and the system interactions of ReSIoT.


	The system architecture of the ReSIoT, which consists of IoT application servers, IoT security domains, GKMS, and AAA system, is shown in Fig.~\ref{fig:SA_system_model}. An IoT security domain is formed by the devices, belonging to the same application. Each device equips with an unique identity associated with the corresponding secret key for the AAA system. IoT application servers support the required capabilities of computing and storage resources, logic operations, and application data and membership managements. 
	Each security domain has one or more dedicated service routers, which are interconnected with IoT devices and can access to global communication networks, such as mobile networks and the Internet.

The service routers are near-user edge IoT devices and considered as SAs, which has sufficient resources to support advanced security algorithms.\footnote{{The SA in the proposed framework is the kernel and different from the SAs in IPSEC and the other related security protocols, which are located at server side. }} 
	For instance, a mobile device (e.g., smartphone) equipping with high computation capability and variant communication interfaces (e.g., UMTS or WLAN) can work as a SA to serve the security of the other resource-limited IoT devices in proximity.
	{Either a user or the manager of an application may deploy the selected SA in advance, and each SA will inherit the security protection mechanism of the underlying communication interface.}
	{By exploiting the computing resources of SA, the computational costs of resource-limited IoT devices can be greatly offloaded.}
	{The deployment of SA brings the following advantages: 
	\begin{enumerate}
		\item reduced complexity of key management for IoT devices;
		\item no requirement of upgrading hardware capabilities of IoT devices to fulfill advanced security mechanisms; and 
		\item high flexibility of adopting new security mechanisms.
\end{enumerate}} 

\subsection{Protocol Stack and System Interactions}\label{subsec:framework_RS}

%
%

\begin{figure}[!t]
	\begin{center}
		\includegraphics[height=7.7cm]{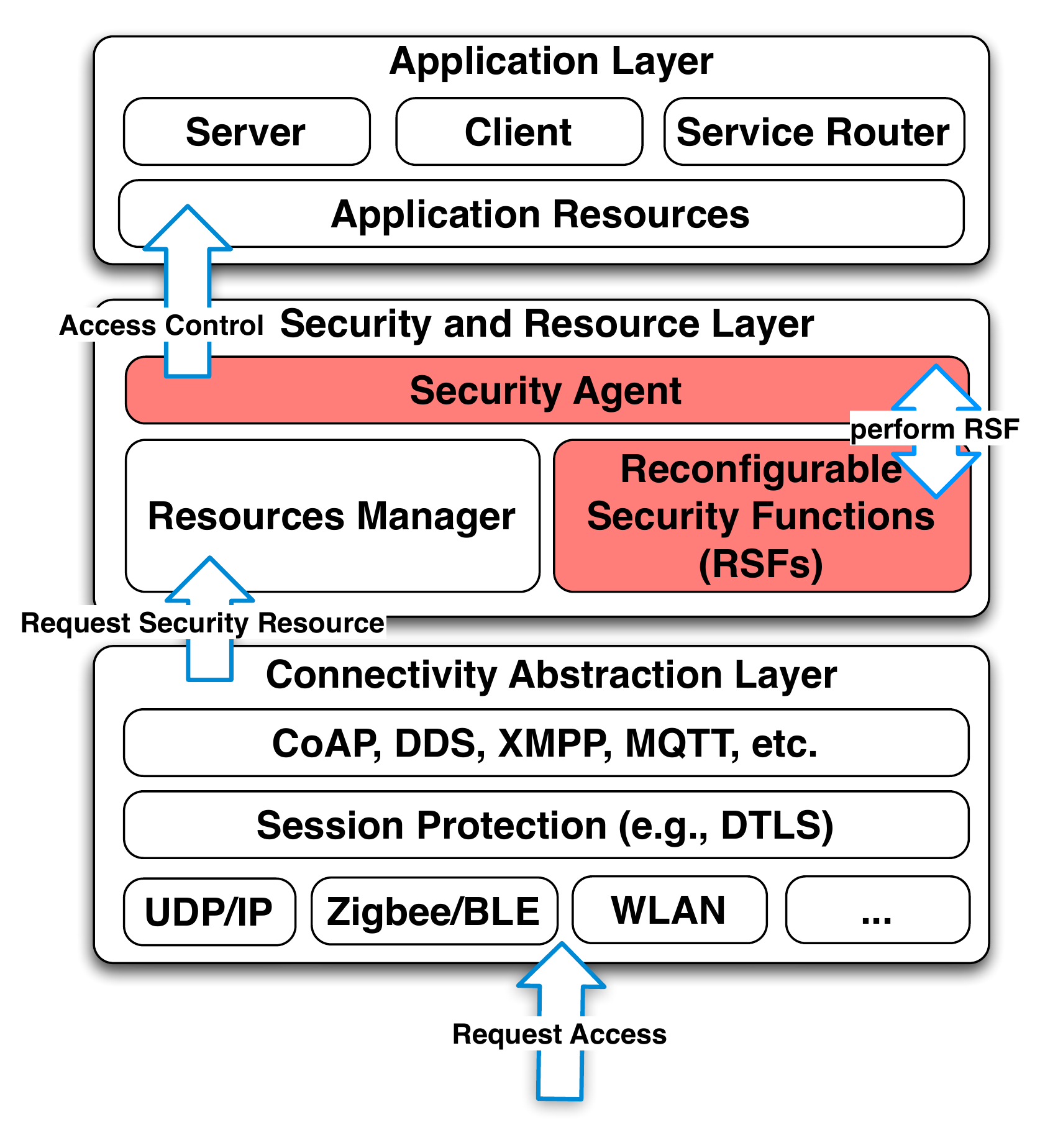} \caption{Protocol Stack of ReSIoT} 
		\label{fig:cross-layer_RS}
	\end{center}
	\vspace{-0.3cm}
\end{figure}

Figure~\ref{fig:cross-layer_RS} 
	shows the protocol stack of ReSIoT, which supports the provision of reconfigurable security for various IoT security requirements. There are three main layers, connectivity abstraction layer, security and resource layer, and application layer.

The connectivity abstraction layer consists of network protocols~(e.g., UDP/ID, Zigbee/bluetooth low energy~(BLE), and WLAN), session protection protocols~(e.g., diagram transport layer security~(DTLS), and TLS), and message-oriented Internet application protocols~(e.g., constrained application protocol~(CoAP), data distribution service~(DDS), extensible messaging and presence protocol~(XMPP), and message queue telemetry transport~(MQTT)). The security and resource layer includes a resource manager, security functions, and security agent. The resource manager maintains the capabilities of computing, communication, and data of IoT devices as resources. The security functions are supported by ReSIoT and also considered as security resources. The security agent is responsible for the functional supports of ReSIoT. The application layer is composed of the applications as resources for servers, clients, and service routers of IoT frameworks.

The RSF is a protocol, which can be realized as a cross-layered middleware, installed on SAs and IoT devices, and it runs on operating system or hardware according to the type of devices. The implementation of RSFs depending on the required operations performed by IoT devices and SA in a complete execution procedure of RSF. Using the RSF, the SA can perform SFs compatibly to IoT devices with all kinds of hardwares, applications, and communication standards.
By the application user interfaces~(APIs) of RSFs, each device is able to support the required security protections for various IoT application requirements. 

As an access request of RSF, sent by a specific application, is received by the connectivity abstraction layer, the resource manager will handle this request with the corresponding SF and perform the RSF with the cooperation of IoT devices and SAs. This enhances the flexibility of supporting various RSFs for not only IoT applications but also communication standards, where communication capabilities are also regarded as resources in IoT frameworks.

\section{Security Protection by ReSIoT}\label{sec:exemplary_IoT_RS}

This section explores how the ReSIoT exploits the advantage of using near-user devices~(i.e., SAs) to perform RSF with IoT devices. First, we discuss how to simplify the key management across multiple IoT applications by ReSIoT. We then introduce the instantiations of two advanced security protection mechanisms, i.e., anonymous authentication and ABAC for data protection, by the construction of ReSIoT. 

\subsection{Key Management and Construction of RSF}\label{subsec:key_management}
\begin{figure} [t]
	\begin{center}
		\includegraphics[height=9.2cm]{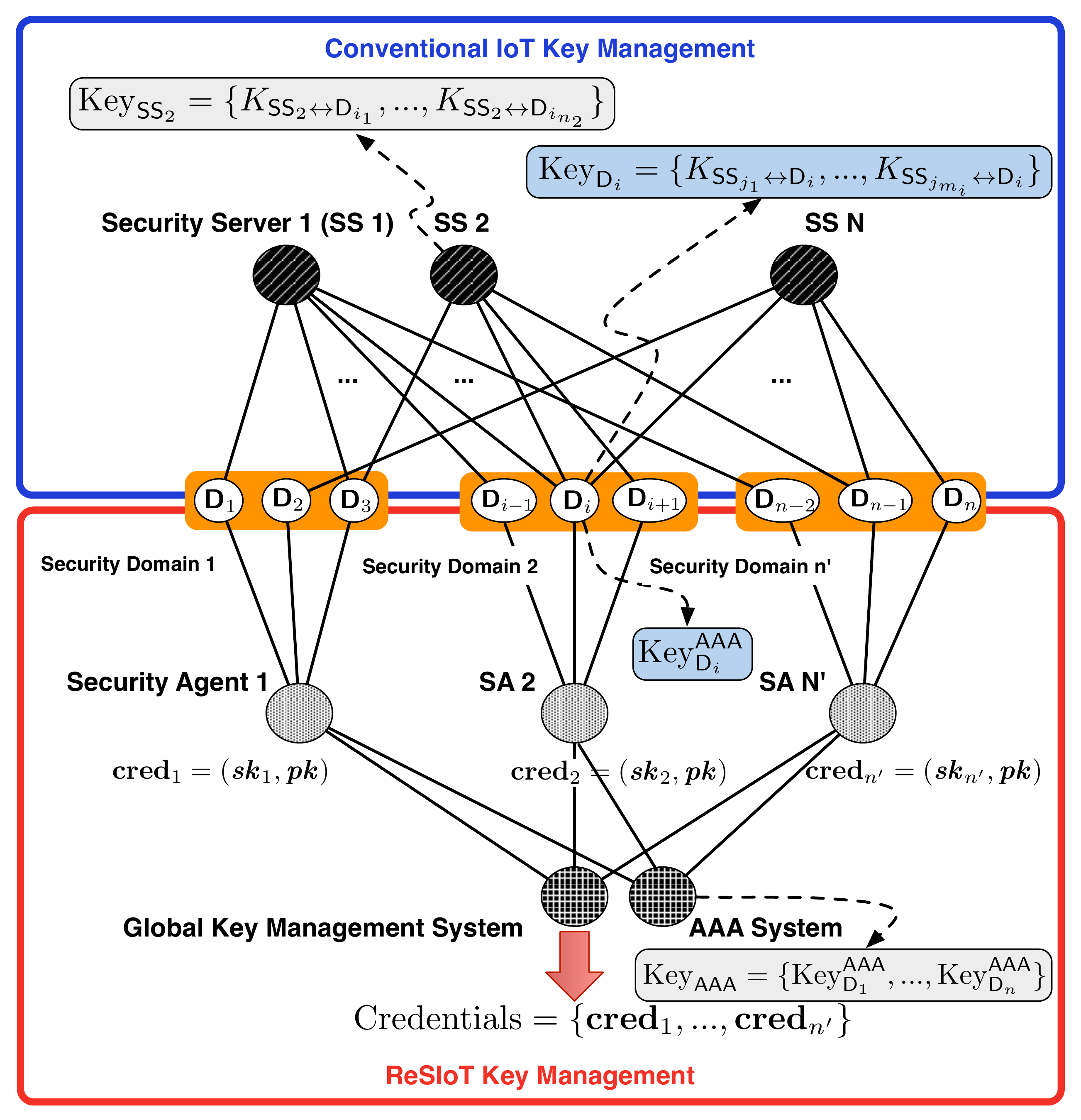} \caption{The key management of IoT in the conventional and ReSIoT architectures.}
		\label{fig:Key_management}
	\end{center}
	\vspace{-0.3cm}
\end{figure}

As depicted in Fig.~\ref{fig:SA_system_model}, each IoT domain deploys its own security keys for the subscriber IoT devices. As shown in Fig.~\ref{fig:Key_management}, {in conventional IoT security, a device $\msf{D}_i$ needs to maintain a set of security keys $\mathrm{key}_{\msf{D}_i}=\{K_{\msf{SS}_{j_1}\leftrightarrow{}\msf{D}_{i}},...,K_{\msf{SS}_{j_{m_i}}\leftrightarrow{}\msf{D}_{i}}\}$ issued by the security servers of different IoT applications.} To reduce the complexity of key management, the IoT reconfigurable security adopts hierarchy-based key management. Specifically, the SA of each security domain maintains a credential for performing each RSFs with the corresponding public/private key pair issued by GKMS. 

Note that a GKMS is considered as the global key management system by trust authorities. The credentials issued by GKMS for certain RSFs, e.g., GS and ABE, should support traceability and revocability~\cite{RGS_LPY12,RABE_ZCDL15}. This offers additional security protections in case of dispute or corruption of SAs. The outputs of RSFs, e.g., signatures and ciphertexts, can be traced to the originator SA. The capability of performing RSFs can be suspended in case of the corruption of SAs.

Each IoT device or thing needs to maintain a secret shared, i.e., $\mathrm{Key}_{\msf{D}_i}^{\msf{AAA}}$, with a global AAA system~(e.g., home subscriber server/authentication center~(HSS/AuC) in mobile networks), provided by network providers. With this secret, the authenticated secure communication among IoT devices and SAs reply on certain AAA-based authentication and key exchange~(AKE) protocol, such as EAP AKE protocol~\cite{EAP_FLH13}. Each SA is assumed to maintain a secure channel to the AAA system. 

\subsection{Construction of RSF} The construction of RSF consists of two parts: \textit{IoT device attachment} and \textit{conversion of SF into RSF}. The attachment procedure, which involves attached IoT device, SA, and AAA system, is to establish mutual authentication among IoT device and SA. One can realize this procedure by a certain AAA-based authentication protocol, i.e., EAP authentication protocol~\cite{EAP_FLH13}, with the shared secret key among IoT devices and AAA system as introduced in the previous section.

The conversion of SF into RSF involves IoT devices and SAs only, where the SAs help to perform SF for IoT devices. Besides the security requirements fulfilled by the specific SF, the design of RSF needs to additionally consider the security requirements, i.e., confidentiality against SA, authenticity against malicious IoT device, and untraceability to IoT devices against SA as introduced in Sec.~\ref{sec:sec_requirements}.
 
\subsection{RSF of Group Signatures for Anonymous Authentication}\label{subsec:anon_auth}
\begin{figure} [t]
	\begin{center}
		\includegraphics[height=7.8cm]{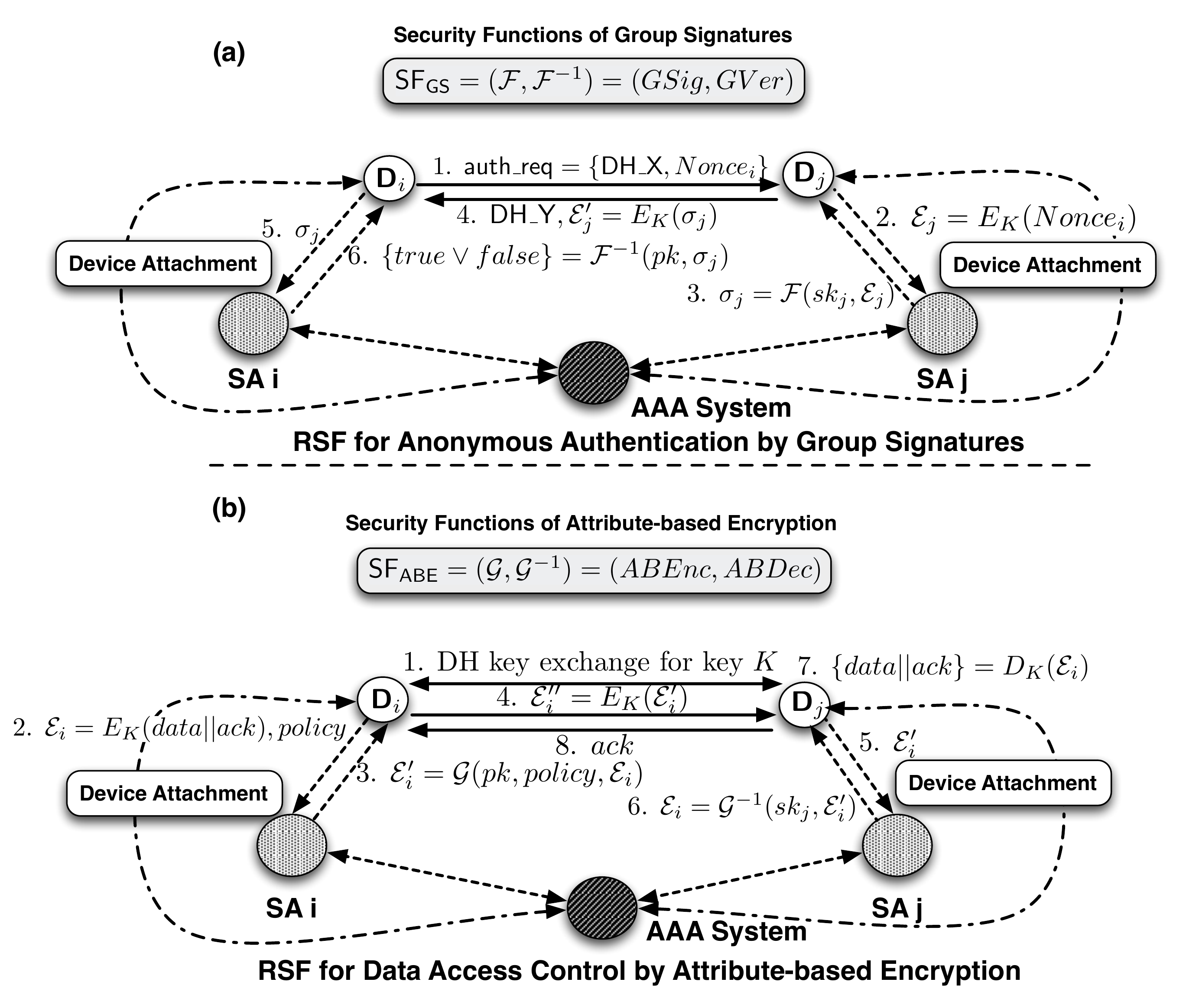} \caption{The constructions of RSF for anonymous authentication and data access control by group signatures and attribute-based encryption, respectively.}
		\label{fig:RSF_Construction}
	\end{center}
	\vspace{-0.3cm}
\end{figure}

By anonymous authentication, one can authenticate an entity without knowing the exact identity, and for this, the identity of every authentication session should be randomized. Group signatures can be utilized to practice anonymous authentication by signing a given message as a signature of the specific group using merely a group-based public-key credential. Figure~\ref{fig:RSF_Construction}(a) shows how to convert a SF of group signatures to its RSF for anonymous authentication between two IoT devices $\mathrmbf{D}_i$ and $\mathrmbf{D}_j$.
The SF consists of two functions $(\mcal{F},\mcal{F}^{-1})=(GSig,GVer)$, where $GSig$ and $GVer$ are the group signing and verifying functions in group signatures, respectively. $GSig$ takes a group private key, $sk_i$, assigned for each SA$_i$ and the signing message as inputs and outputs a group signature, $\sigma_i$, which can be verified by $GVer$ with the group public key, $pk$, which outputs \textsf{true} or \textsf{false} as the result of verifying $\sigma_i$.

Suppose that two devices attached to individual SAs, i.e., $\msf{SA}_i$ and $\msf{SA}_j$, by an EAP-based authentication with the global AAA system for the attachment. 
After the attachment, the instantiation of RSF by group signature SF consists of the following procedure: 
1) $\mathrmbf{D}_i$ sends $\msf{auth\_req}=\{\msf{DH\_X},Nonce_i\}$ to authenticate $\mathrmbf{D}_j$, where $\msf{DH\_X}=g^{x}$ is one of the Diffie-Hellman key exchange tuple and $Nonce_i$ is a random number, 
2) $\mathrmbf{D}_j$ generates a shared key $K=g^{xy}$ by its Diffie-Hellman tuple $\msf{DH\_Y}=g^{y}$ and the received $\msf{DH\_X}$, encrypts $Nonce_i$ as $\mcal{E}_j=E_{K}(Nonce_i)$ with a symmetric encryption, and sends $rand$ to $\msf{SA}_j$, 
3) $\msf{SA}_j$ generates a group signature $\sigma_j$ on $\mcal{E}_j$ with $sk_j$ and sends $\mcal{E}_j$ to $\mathrmbf{D}_j$, 
4) $\mathrmbf{D}_j$ encrypts $\sigma_j$ with the shared key $K$ as $\mcal{E}'_j=E_{K}(\sigma_j)$, and sends $(\msf{DH}\_Y,\mcal{E}'_j)$ to $\mathrmbf{D}_i$,
5) $\mathrmbf{D}_i$ decrypts $\mcal{E}'_j$ to obtain $\sigma_j$ with $K$ computed by the received $\msf{DH_Y}$ and sends $\sigma_j$ to SA$_i$ for the verification of $\sigma_j$, and 
6) SA$_j$ verifies $\sigma_j$ by performing $\mcal{F}^{-1}(pk,\sigma_j)$ and return $true$ or $false$ as the result if the group signature from $\mathrmbf{D}_j$ is correct or not. If so, the anonymous authentication is done among $\mathrmbf{D}_i$ and $\mathrmbf{D}_j$ by the RSF of group signatures.


\subsection{RSF of Attribute-based Encryption for Attribute-based Access Control}
The ABAC allows a device to access protected things~(e.g., multimedia content) or devices~(e.g., sensors) with the verification of the attributes assigned to IoT devices and the policy set for the protected target.
The ABE provides secure ABAC by encrypting messages with the public key of ABE and a specified policy (i.e., a {predicate} of attributes). Only the devices with legitimate secret keys associated with the attributes, which satisfy the policy, can decrypt the encryption. This implies the verification of attributes and policies is achieved via the attribute-based encryption and decryption of ABE.
Here, we want to build a RSF of ABE by the SF of ABE to practice the fine-grained access control in IoT. The SF consists of two functions, $(\mcal{G},\mcal{G}^{-1})=(ABEnc,ABDec)$, where $ABEnc$ and $ABDec$ are the encryption and decryption functions of ABE.

Figure~\ref{fig:RSF_Construction}(b) depicts how to convert a ABE SF to the its RSF for secure ABAC among two IoT devices with SAs.
The construction follows the procedures below: 
1) In the beginning, both $\mathrmbf{D}_i$ and $\mathrmbf{D}_j$ perform the attachment procedures, which are the same as that in the RSF of group signatures, and exchange a shared key $K$ by Diffie-Hellman key exchange;
2) $\mathrmbf{D}_i$ encrypts its data and acknowledgement $ack$ with $K$ as $\mcal{E}_i=E_{K}(data||ack)$, and sends $\mcal{E}_i$ and the ABAC policy to SA$_i$;
3) $SA_i$ encrypts $\mcal{E}_i$ by $\mcal{G}$ with the public key $pk$ and $policy$ as $\mcal{E}'_i$ and sends back to $\mathrmbf{D}_i$;
4) $\mathrmbf{D}_i$ encrypts $\mcal{E}'_i$ with $K$ as $\mcal{E}''_i=E_{K}(\mcal{E}'_i)$ and sends $\mcal{E}'_i$ to $\mathrmbf{D}_j$;
5) $\mathrmbf{D}_j$ decrypts $\mcal{E}''_i$ to obtain $\mcal{E}'_i$ and forwards to $\msf{SA}_j$;
6) $\msf{SA}_j$ decrypts $\mcal{E}'_i$ by $\mcal{G}^{-1}$ with its private key $sk_j$ to obtain $\mcal{E}'_i$ and sends it back to $\mathrmbf{D}_j$, and 
7) $\mathrmbf{D}_j$ decrypts $\mcal{E}'_i$ to obtain $data$ and $ack$, and sends $ack$ to confirm the correctness of the decryption.
If the attributes of $sk_j$ satisfy $policy$, $\mathrmbf{D}_j$ will obtain the exact same $data$ and $ack$ shared with $\mathrmbf{D}_i$.

\subsection{Security Analysis}
We discuss how the two proposed RSFs can fulfill the three additional security requirements, presented in Sec.~\ref{sec:sec_requirements}, as follows. First, since a session key is shared by Diffie-Hellman key agreement in every RSF session, the SA can only compute encrypted messages and learn no messages exchanged between two IoT devices from computing specified SF for IoT devices. This guarantees the confidentiality against honest-but-curious SAs. Second, IoT devices fulfill the authenticity of every IoT device attaching to the SA in its proximity. Finally, the communication sessions, launched by the same IoT device will not be traceable when anonymous authentication is adopted in the IoT device attachment procedure. Hence, the computing of SFs on SAs will not break the security of communications among IoT devices.

\section{Performance of ReSIoT}

This section evaluates the communication and computation costs of two RSFs, described in Fig.~\ref{fig:RSF_Construction}. The advanced cryptographic algorithms are used such as Boneh-Boyen-Shacham group signatures~(BBS GS)~\cite{GSig_BBS04} and Goyal-Sahai-Pandey-Waters attribute-based encryption~(GSPW ABE)~\cite{AttributeEnc_GSPW06} for anonymous authentication and ABAC. The SA is commonly considered as a relatively powerful device among IoT devices. Hence, we take a desktop computer (Apple Macbook Air 2012 model equipped with Intel Core i5~(dual-core) 1.8 GHz CPU and 4 GB RAM) as a SA and a smart phone (ASUS Zenfone 2 ZE551ML equipped with Quad-core 1.8 GHz CPU and 2 GB RAM model) as IoT devices to observe the performance enhancement with the ReSIoT in the experiments.{\footnote{{These devices have been adopted with the expectation of the computing capability evolution, which will be made at even small IoT devices in the future.}}}

\subsection{Comparison of Computational Cost for Cryptographic Algorithms}
Before comparing the performance of reconfigurable security with legacy security solutions, we first analyze the computational costs of two cryptographic algorithms, BBS GS and GSPW ABE on the testbed platforms, respectively. The algorithms employ the bilinear pairing operation~\cite{Pairing_BLS01}, which is an elliptic curve cryptography~(ECC) arithmetic operation and needs considerable computational cost to {execute the algorithms}. In Table~\ref{table:computation_cost_crypto_alg}, we present the computational time of the bilinear pairing related operations, including multiplication~(Mul), exponentiation~(Power), and pairing in the same bilinear pairing groups on both platforms by Java Pairing-based Cryptography~(JPBC) library. Using those computational times, we can obtain the computational times of two SFs on SA and IoT device, respectively.

\begin{table*}[t!] 
\caption{Computation Costs of Cryptographic Algorithms}
\label{table:computation_cost_crypto_alg}
\begin{center}
\begin{tabular}{|M{3cm}||M{2.3cm}|M{1.2cm}|M{1.2cm}|M{1.3cm}|M{1.3cm}|M{1.3cm}|M{1.3cm}|} \hline
 & Pairing, $T_{p}$~($\mbb{G}_1 \!\times \! \mbb{G}_2 \! \rightarrow \! \mbb{G}_T$) & Power, $T^{\mbb{G}_1}_{\msf{exp}}$~($\mbb{G}_1$)& Mul, $T_{mul}^{\mbb{G}_1}$~($\mbb{G}_1$)  & Power, $T^{\mbb{G}_2}_{\msf{exp}}$~($\mbb{G}_2$)& Mul, $T_{mul}^{\mbb{G}_2}$~($\mbb{G}_2$)& Power, $T^{\mbb{G}_T}_{\msf{exp}}$~($\mbb{G}_T$)& Mul, $T_{mul}^{\mbb{G}_T}$~($\mbb{G}_T$)\\ \hline
IoT device~(Smart phone)& 271.0 ms& 122.7 ms& 123.2 & 121.8 ms & 115.3 ms & 40.8 ms& 39.5 ms \\ \hline
Security agent~(PC)   & 14.8 ms & 13.6 ms & 12.8 ms & 12.6 ms & 12.6 & 1.1 ms & 1.1 ms\\ \hline\hline
&\multicolumn{3}{c|}{\pbox{5.4cm}{\vspace*{0.1cm}BBS Group Signature$^{(1)}$\vspace*{0.1cm}}}&\multicolumn{4}{c|}{\pbox{4.8cm}{\vspace*{0.1cm}GSPW Attribute-based Encryption\\(for 50 attributes)$^{(2)}$\vspace*{0.1cm}}}\\ \hline
&\multicolumn{1}{c}{Signing}&\multicolumn{2}{c|}{Verifying}&\multicolumn{2}{c}{Encrypt}&\multicolumn{2}{c|}{Decrypt}\\ \hline
Computation time at IoT device for SF, $t_{SF}^{\msf{D}}$ & \multicolumn{1}{c}{2409.3 ms} & \multicolumn{2}{c|}{1786.8 ms} & \multicolumn{2}{c}{6380.9 ms} & \multicolumn{2}{c|}{1863.0 ms} \\ \hline
Computation time at SA for SF, $t_{SF}^{\msf{SA}}$   & \multicolumn{1}{c}{208.5 ms} & \multicolumn{2}{c|}{224.7 ms} & \multicolumn{2}{c}{706.4 ms} & \multicolumn{2}{c|}{95.4 ms} \\ \hline
Processing time of conventional SF, $T_{SF}$   & \multicolumn{1}{c}{2465.3 ms} & \multicolumn{2}{c|}{1842.8 ms} & \multicolumn{2}{c}{6436.9 ms} & \multicolumn{2}{c|}{1919 ms} \\ \hline
Processing time of RSF, $T_{RSF}$   & \multicolumn{1}{c}{509.837 ms} & \multicolumn{2}{c|}{526.037 ms} & \multicolumn{2}{c}{1067.737 ms} & \multicolumn{2}{c|}{396.737 ms} \\ \hline
Reduced Processing Time by RSF & \multicolumn{1}{c}{79.32\%}& \multicolumn{2}{c|}{71.46\%}&\multicolumn{2}{c}{83.42\%} &\multicolumn{2}{c|}{79.33\%}\\ \hline
\multicolumn{8}{|l|}{\pbox{15.5cm}{\vspace*{0.1cm}
(1) Computation time of BBS Group Signature: (Signing) $9T^{\mbb{G}_1}_{\msf{exp}}$+$3T_{mul}^{\mbb{G}_1}$+$3T_{\msf{exp}}^{\mbb{G}_T}$+$3T_{p}$, (Verifying) $8T_{\msf{exp}}^{\mbb{G}_1}+4T_{mul}^{\mbb{G}_1}+5T_{\msf{exp}}^{\mbb{G}_1}+4T_{p}$}} \\
\multicolumn{8}{|l|}{\pbox{15.5cm}{\vspace*{0.1cm}
(2) Computation time of Attribute-based Encryption: (Encrypt) $(N_{a}+1)T_{\msf{exp}}^{\mbb{G}_1}+T_{mul}^{\mbb{G}_1}$ for $N_a$ number of attributes, (Decrypt) \hspace*{2mm} $\lceil\log{N_{a}}\rceil(T_{p}+T_{mul}^{\mbb{G}_{T}})$}} \\
\multicolumn{8}{|l|}{\pbox{15.5cm}{\vspace*{0.1cm}
- $t_{SF}^{\msf{D}}$ and $t_{SF}^{\msf{SA}}$ are the computation times of the SF with respect to the specified cryptographic algorithm on smart phone and PC,
\hspace*{2mm}respectively.}} \\
\multicolumn{8}{|l|}{\pbox{15.5cm}{\vspace*{0.1cm} - $t_{RSF}^{\msf{D}}=t_{DH}^{\msf{D}}+t_{\msf{Enc}}^{\msf{D}}$ is the computation time required for RSF on smart phone, where  $t_{DH}^{\msf{D}}$ = 2.31 ms is the computation time of \hspace*{2mm}Diffie-Hellman key agreement with key size of 1024-bit and $t_{\msf{Enc}}^{\msf{D}}$ = 0.027 ms is the computation time of AES encrytion/decryption \hspace*{2mm}with input of 160-bit on smart phone.}}\\ \hline
\end{tabular}
\end{center}
\vspace{-0.3cm}
\end{table*}

\subsection{Processing Time}
We compare the performance of RSFs with the legacy solutions for two security requirements, i.e., anonymous authentication and ABAC. 
Here, the legacy solutions mean the ones simply perform SFs~(i.e., BBS GS and GSPW ABE) on IoT devices for the aforementioned security requirements. First, the processing time for completing RSF with IoT devices and SA can be denoted as $T_{RSF}=t_{attach}^{\msf{D}}+t_{RSF}^{\msf{D}}+t_{RSF}^{\msf{SA}}~(\approx t_{SF}^{\msf{SA}})+t^{\msf{D}}_{COM}+t^{\msf{D}-\msf{SA}}_{COM}$, where $t_{attach}^{\msf{D}}$ is the attachment time of IoT devices to SAs, $t_{RSF}^{\msf{D}}$ is the computation time of IoT devices for RSF, $t_{RSF}^{\msf{SA}}$ is the computation time of RSF on SAs and it is basically equal to the computation time of SF on SAs, $t_{SF}^{\msf{SA}}$, $t_{COM}^{\msf{D}}$ is the communication time between IoT devices, and $t_{COM}^{\msf{D}-\msf{SA}}$ is the communication time between IoT devices and SAs. Since each device only needs to attach the SA in its communication coverage at the beginning, we therefore eliminate the time of attachment to SA in evaluating $T_{RSF}$. We denote the processing time of completing SF with merely IoT devices as $T_{SF}=t_{SF}^{\msf{D}}+t_{COM}^{\msf{D}}$, where $t_{SF}^{
\msf{D}}$ is the computation time of SF on IoT device.
We estimate $t_{COM}^{\msf{D}}$ by observing the communication of WiFi direct, which is one of wireless communication standards for device-to-device communications and $t_{COM}^{\msf{D}-\msf{SA}}$ by observing the communication between smartphone and a PC server via wireless AP. In the experiment, $t_{COM}^{\msf{D}}$ is 56 ms and $t_{COM}^{\msf{D}-\msf{SA}}$ is 243 ms in average by transmitting 1024-bit string for 1000 times.

{The computation and communication time of the signing and verification of BBS GS and the encryption and decryption of GSPW ABE are summarized in Table~\ref{table:computation_cost_crypto_alg}.}
{From Table~\ref{table:computation_cost_crypto_alg}, we can see that the processing time has been reduced by from 83.42\% to 71.46\% by the RSF.}
This result shows even though the RSFs require additional communications among IoT devices and SAs, the processing times of RSFs remain to outperform those of the conventional SFs thanks to the significant amount of high complexity computations being offloaded to SAs.

{We then can estimate the rate of completing functions required by a security request within the request expiration time, which we call it as the success rate of security requests.}
%
{Here, we consider the load at the SA as well. 
A SA may serve multiple devices at once, and the SA will reserve its resource~(e.g., CPU and required memory for processing) for each incoming security request issued by devices or things to serve in a first-come-first-serve~(FCFS) manner.}
The incoming request needs to wait in a queue to be processed when the SA is busy for other requests. The security request will fail whenever the device or thing issuing request loses the connection to the SA or 
the session time of the request is expired before completing the process for the security request.
Hence, the expected processing time of each security service can be determined by averaging out the processing time $t_{RSF}^{\msf{SA}}$ when $t_{\msf{RSF}}^{\msf{SA}} \leq t_{\msf{exp}}$, where $t_{\msf{exp}}$ is the expiration time of a request, $t_{\msf{RSF}}^{\msf{SA}}=t^{\msf{SA}}_{Q} + t_{SF}^{\msf{SA}}$, and $t_{Q}^{\msf{SA}}$ is the waiting time in queue for a request.

\begin{figure} [t!]
	\begin{center}
		\includegraphics[height=6.8cm]{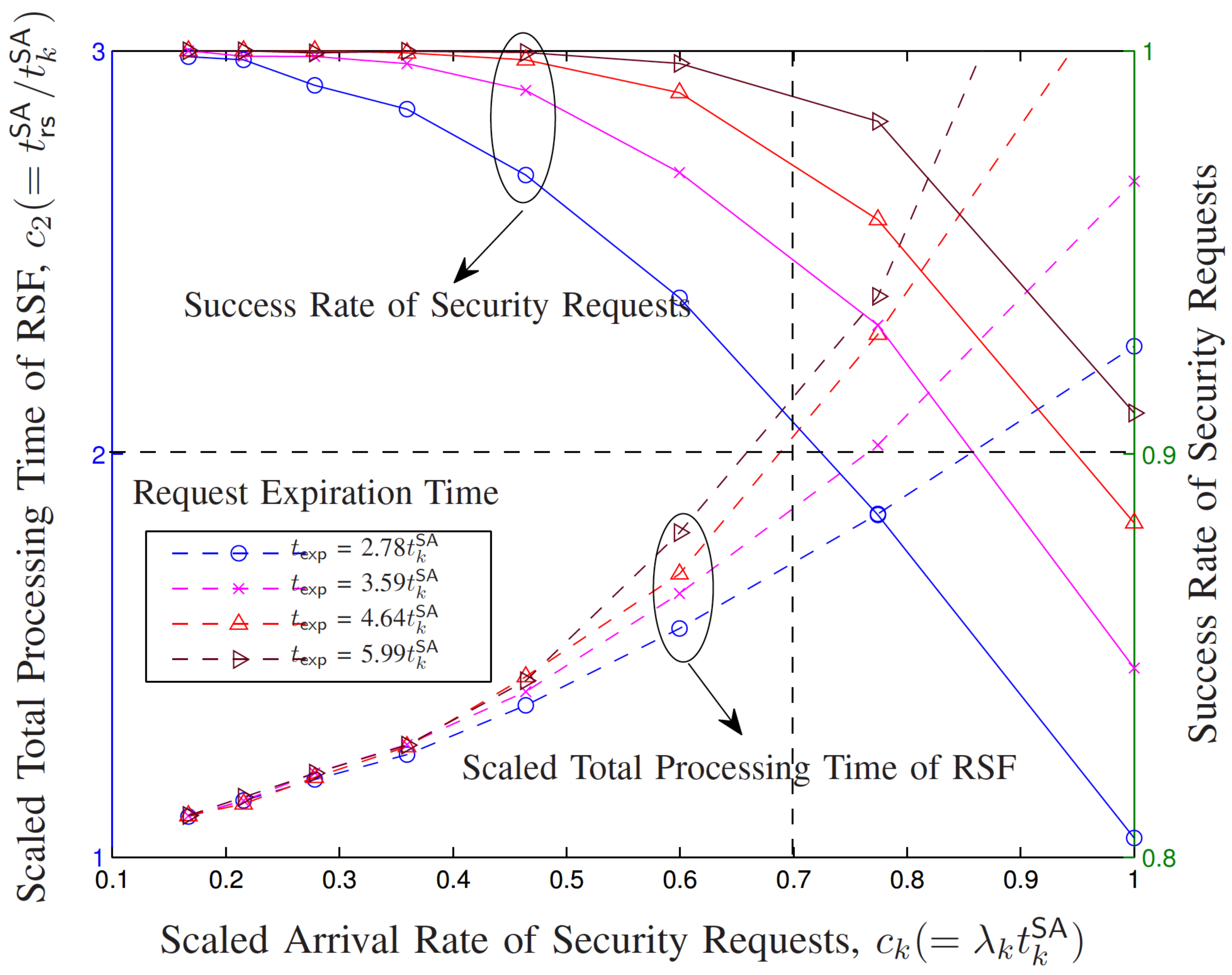} \caption{This figure evaluates the system processing time and success rate of security requests with different arrival rate of security requests and different request expiration time.}
		\label{fig:simulation}
	\end{center}
	\vspace{-0.3cm}
\end{figure}

{Figure~\ref{fig:simulation} shows the total processing time and the success rate of security requires according to the arrival rate of security requests for different values of expiration time of security request $t_\msf{exp}$. In this figure, $c_k (= \lambda_{k} t_{SF}^{\msf{SA}})$ is the scaled arrival rate of security requests~(i.e., the arrival rate in $t_{\msf{SF}}^{\msf{SA}}$), and $c_2 (= t_{\msf{RSF}}^{\msf{SA}}/t_{SF}^{\msf{SA}})$ is scaled total processing time of RSF.}
 %
 Here, we use a fixed value, $208.5$~ms, for $t_{SF}^{\msf{SA}}$ (the computational time of a signing operation of BBS group signature) and increase {$\lambda_{k}$ up to $\lambda_{k} = 1/t_{SF}^{\msf{SA}}$} (i.e., $c_k = 1$)
 to keep a steady state of queue in the simulation.
From Fig.~\ref{fig:simulation}, we can first see that even for $c_k = 1$, the success rate is greater than $80\%$. For the arrival rate less than $0.7t_{k}^{\msf{SA}}$ (i.e., $c_k < 0.7$), the success rate is higher than $90\%$ for all cases of request expiration time. 

\vspace{-0.2cm}

\section{Summary}\label{sec:conclusion}

In this article, after overviewing issues and existing solutions of IoT security, 
we introduce a reconfigurable security framework based on edge computing,
which utilizes a near-user edge devices, i.e., a SA, with stronger computation capability for IoT security.
Based on the framework, one can design a reconfigurable security function protocol, interacting with SA, to resolve any specific security requirements. We also provide two exemplary reconfigurable security function protocols for anonymous authentication and secure attribute-based access control.
Through the performance analysis, it is shown that {reconfigurable security function protocols outperform} legacy solutions even though additional communications required to interact with SA.
{The reconfigurable security can provide security protections for IoT devices {with better flexibility and scalability in hardwares (e.g., computational capability and memory) and softwares (e.g., various of standards and the complexity of key management due to comprehensive application scenarios).}} 

\vspace{-0.3cm}
\bibliographystyle{IEEEtran}
\bibliography{RS_IoT_Security}  

\end{document}